\documentclass[3p, table, xcdraw]{elsarticle}

\usepackage[utf8]{inputenc}
\usepackage{graphicx}
\usepackage{amsmath}
\usepackage{amssymb}
\usepackage{verbatim}
\usepackage[font=scriptsize]{caption}
\usepackage{subcaption}
\usepackage{makecell}
\usepackage{float}
\usepackage{scalerel}
\usepackage{microtype} 
\usepackage[english]{babel}
\usepackage{tikz}
\usepackage{orcidlink}
\usepackage{hyperref}
\usepackage{blindtext}
\usepackage{booktabs}
\usepackage{adjustbox}
\usepackage{filecontents}

\usetikzlibrary{svg.path}
\hypersetup{pdfauthor={Name}}
\graphicspath{{./images/}}
\makeatletter
\def\ps@pprintTitle{
 \let\@oddhead\@empty
 \let\@evenhead\@empty
 \def\@oddfoot{}
 \let\@evenfoot\@oddfoot}
\makeatother

\begin{document}
{\hypersetup{hidelinks}
\begin{frontmatter}

\title{\LARGE Predicting Hydroxyl Mediated Nucleophilic Degradation and Molecular Stability of RNA Sequences through the Application of Deep Learning Methods} 

\author[1]{Ankit Singhal\corref{cor1}\,\orcidlink{0000-0001-5563-6493}}
 \ead{ankit.singhal@isbasel.ch}
 \cortext[cor1]{Corresponding author}
 \affiliation[1]{organization={International School Basel}, addressline={Fleischbachstrasse 2}, postcode={4153}, city={Reinach, Baselland}, country={Switzerland}}
 
\begin{abstract}
	Synthesis and efficient implementation mRNA strands has been shown to have wide utility, especially recently in the development of COVID vaccines. However, the intrinsic chemical stability of mRNA poses a challenge due to the presence of 2’-hydroxyl groups in ribose sugars. The -OH group in the backbone structure enables a base-catalyzed nucleophilic attack by the deprotonated hydroxyl on the adjacent phosphorous and consequent self-hydrolysis of the phosphodiester bond. As expected for in-line hydrolytic cleavage reactions, the chemical stability of mRNA strands is highly dependent on external environmental factors, e.g. pH, temperature, oxidizers, etc. Predicting this chemical instability using a computational model will reduce the number of sequences synthesized and tested through identifying the most promising candidates, aiding the development of mRNA related therapies. This paper proposes and evaluates three deep learning models (Long Short Term Memory, Gated Recurrent Unit, and Graph Convolutional Networks) as methods to predict the reactivity and risk of degradation of mRNA sequences. The Stanford Open Vaccine dataset of 6034 mRNA sequences was used in this study. The training set consisted of 3029 of these sequences (length of 107 nucleotide bases) while the testing dataset consisted of 3005 sequences (length of 130 nucleotide bases), in structured (Lowest Entropy Base Pair Probability Matrix) and unstructured (Nodes and Edges) forms. The stability of mRNA strands was accurately generated, with the Graph Convolutional Network being the best predictor of reactivity ($RMSE = 0.249$) while the Gated Recurrent Unit Network was the best at predicting risks of degradation ($RMSE = 0.266$). Combining all target variables, the GRU performed the best with $76\%$ accuracy. Results suggest these models can be applied to understand and predict the chemical stability of mRNA in the near future.
\end{abstract}
\begin{keyword}
	mRNA Stability \sep Nucleotide Sequence Prediction \sep Recurrent Neural Network \sep Graph Convolutional Network \sep Gated Recurrent Unit \sep Long Short Term Memory
\end{keyword}
\end{frontmatter}}

\section{Introduction}
\begin{figure}[H]
    \centering
    \includegraphics[scale=0.4]{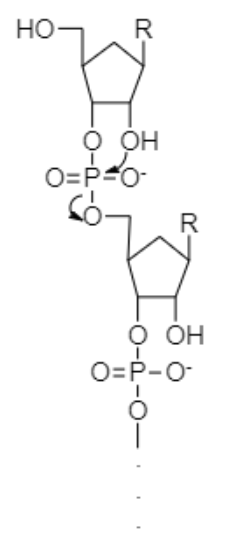}
    \caption{RNA hydrolysis via nucleophilic attack by 2'-hydroxyl group on phosphodiester bond in sugar phosphate backbone (R = nucleotide base).}
    \label{fig:rna_structure}
\end{figure}

\hspace{\parindent}Over the last two decades, there has been increasing interest in the field of RNA-based technologies in the creation of prophylactic vaccines, among other applications such as cancer immunotherapies, induction of pluripotent stem cells, and \textit{in vivo} delivery of mRNA to replace or supplement proteins. They are widely regarded to be plausible alternatives to conventional approaches due to their high potency, capacity for quick and low-cost manufacturing, and relatively safe administration. They are currently being researched for many diseases, with the Pfizer-BioNTech vaccine against SARS-CoV-2 being the first to be approved for human use \cite{pardi, zhang, pfizer}. One of the primary barriers in the creation of such therapeutics is the fragility of the RNA molecule; they are susceptible to rapid degradation within minutes to hours \cite{scope} and as such need to be freeze dried or incubated at low temperatures in order to be kept stable. The primary mechanism of this degradation is through RNA hydrolysis, in which the phosphodiester bond in the sugar-phosphate bond is broken as a result of a base-catalyzed nucleophilic attack by the 2'-hydroxyl group (Refer to Figure \ref{fig:rna_structure}).

In the current pandemic, vaccines are seen as the most promising means to control the novel coronavirus, but with current limitations on mRNA vaccines, efficient \textit{in vivo} delivery of such mRNA molecules seems improbable; it would likely only reach a fraction of the population, mostly relegated to countries with a higher level of infrastructural development \cite{swift}.

Therefore, research into the the stability and degradation of RNA molecules has received continued interest, to date largely consisting of traditional statistical approaches and biophysical models. However, it still remains unclear exactly which parts of RNA molecules are more prone to spontaneous degradation and thus difficult to accurately predict the reactivity and degradation of mRNA \cite{wayment}. Therefore, experimentation, an incredibly time consuming process, is the default method in determining these values.

The aim of this manuscript is to present three possible Deep Learning approaches to this problem through the usage of the Stanford OpenVaccine dataset. Two variants of Recurrent Neural Networks (RNNs) are employed, Long Short Term Memory Networks (LSTMs) and Gated Recurrent Unit Networks (GRUs), along with a variant of a Graph Neural Network (GNN), the Graph Convolutional Network (GCN). These models are applied and compared to assess whether Machine Learning methods can provide helpful results in predicting the reactivity and degradation of mRNA molecules. This can save significant resources in the development of mRNA vaccines by helping identify the most promsing candidates using in silico methods.
\section{Materials and Methods}
\subsection{Materials}
\subsubsection{Software/Packages Used}
The majority of this work was done in the Python programming language using a TensorFlow back end with a surface level Keras API. Refer to Table  ~\ref{tab:packages} for all the software/packages used throughout the creation and testing of the models along with related information \cite{python, tensorflow, keras, sklearn, pandas, numpy, matplotlib, viennarna, contrafold}.
\begin{table*}[ht]
\scriptsize
\centering
\begin{tabular}{lll}
\Xhline{3\arrayrulewidth}
Software/Package & Use                                                                                                    & Developer                  \\ \Xhline{3\arrayrulewidth}
Python 3.7       & \begin{tabular}[c]{@{}l@{}}Used to write the code for the models \\ discussed.\end{tabular}            & Python Software Foundation \\
TensorFlow       & \begin{tabular}[c]{@{}l@{}}Used as a backend for the majority of \\ the models presented.\end{tabular} & Google                     \\
Keras            & \begin{tabular}[c]{@{}l@{}}Used as a high level API for some \\ parts of the model.\end{tabular}       & Google                     \\
SKLearn          & Used for k-Fold Cross Validation.                                                                      & Cornapeau and Matthieu     \\
Pandas           & Used for data handling.                                                                                & McKinney                   \\
Numpy            & Used for data handling.                                                                                & Oliphant                   \\
Matplotlib       & Used to create figures in manuscript.                                                                  & Droettboom and Caswell     \\
ARNiE            & Used to generate augmentation data.                                                                    & DAS Labs                   \\ \Xhline{3\arrayrulewidth}
\end{tabular}
\caption{All the packages/software used in the creation of the models presented.}
\label{tab:packages}
\end{table*}

\subsubsection{Dataset Description}
\hspace{\parindent}   The dataset used in this manuscript to evaluate the models is called the ``Stanford OpenVaccine" dataset \cite{kaggle}. It consists of 6034 RNA sequences. The training/validation set consisted of 3029 of these sequences with a length of 107 nucleotide bases. The testing dataset consists of 3005 sequences with a length of 130 nucleotide bases. Due to experimental limitations (in collecting the data according to the researchers at Stanford \cite{kaggle}), measurements on the final bases in the sequences cannot be carried out, therefore, only 68 (for the sequences with length 107) and 91 (for the sequences with length 130) of the first bases in the respective sequences have experimental data associated with them.

Three predictors were associated with each sequence: the sequence itself (described in A,G,C, and U bases), the expected structure of the molecule, and the Predicted Loop Type (derived from a simulation of the secondary structure of the RNA molecule). A Base Pair Probability Matrix was also provided for each individual sequence indicating the probability of certain base-pair interactions. Five experimentally-determined sets of values (henceforth referred to as 'target values') were also given for the first 68 or 91 base pairs in the sequence: Reactivity values, degradation values at pH 10, degradation values at pH 10 with added Magnesium, degradation values at 50$^\circ$, and degradation values at 50$^\circ$ with added Magnesium. Refer to Table ~\ref{tab:description} for more information.

\begin{table*}[ht]
\centering
\tiny
\begin{tabular}{llll}
\Xhline{3\arrayrulewidth}
Feature                                                        & Classification  & Description                                                                                                                                                                                                                                                   & Sample              \\ \Xhline{3\arrayrulewidth}
Sequence                                                       & \textit{Input}  & \begin{tabular}[c]{@{}l@{}}A sequence of 107 letters corresponding\\ to the four bases in the sequence.\end{tabular}                                                                                                                                          & A, G, U, U, C, ...  \\
Structure                                                      & \textit{Input}  & \begin{tabular}[c]{@{}l@{}}Expected structure of the molecule \\ (length = 107). '(' and ')' refer to a base \\ pair interaction. All '.' in the middle are \\ associated with no BP interactions.\end{tabular}                                               & (..()...)()(... ... \\
\begin{tabular}[c]{@{}l@{}}Predicted \\ loop type\end{tabular} & \textit{Input}  & \begin{tabular}[c]{@{}l@{}}Predicted secondary structure of the \\ RNA molecule at different points. 'S' \\ refers to a stem structure, 'M' multiloop,\\ 'I' internal loop, 'B' bulge, 'H' hairpin \\ loop, 'E' dangling end, 'X' external loop.\end{tabular} & S, S, M, S, H, ...  \\
Reactivity                                                     & \textit{Target} & \begin{tabular}[c]{@{}l@{}}Reactivity values at each individual \\ point in thesequence.\end{tabular}                                                                                                                                                         & 1.23, 3.46, ...     \\
Deg pH 10                                                      & \textit{Target} & Degradation values at pH 10.                                                                                                                                                                                                                                  & 0.89, 2.44, ...     \\
Deg pH 10 Mg                                                   & \textit{Target} & \begin{tabular}[c]{@{}l@{}}Degradation values at pH 10 with \\ added Mg.\end{tabular}                                                                                                                                                                         & 1.28, 0.88, ...     \\
Deg 50$^\circ$C                                                & \textit{Target} & Degradation values at 50$^\circ$C.                                                                                                                                                                                                                            & 2.02, 1.87, ...     \\
Deg 50$^\circ$C Mg                                             & \textit{Target} & \begin{tabular}[c]{@{}l@{}}Degradation values at 50$^\circ$C \\ with added Mg.\end{tabular}                                                                                                                                                                   & 1.11, 2.44, ...     \\ \Xhline{3\arrayrulewidth}
\end{tabular}
\caption{Inputs and target features of the Stanford OpenVaccine dataset.}
\label{tab:description}
\end{table*}

The task of the algorithms that are presented in this paper is to take the sequence and other structural features of an RNA molecule (features marked as `inputs') and predict its stability (through the five target values).
\subsubsection{Data Representation}
\begin{figure}[H]
    \centering
    \includegraphics[scale=0.8]{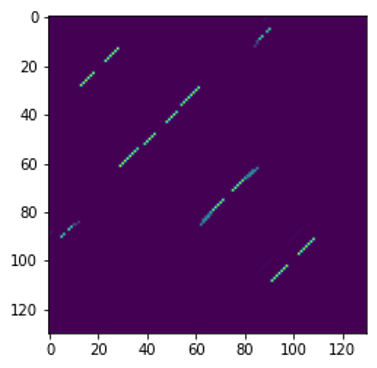}
    \caption{Visualization of a sample BPPM Matrix. Purple indicates no base pair interaction, the more green a color, the higher the interaction between those two bases (produced by author).}
    \label{fig:bppm}
\end{figure}

\begin{figure*}[ht]
    \centering
    \includegraphics[scale=0.4]{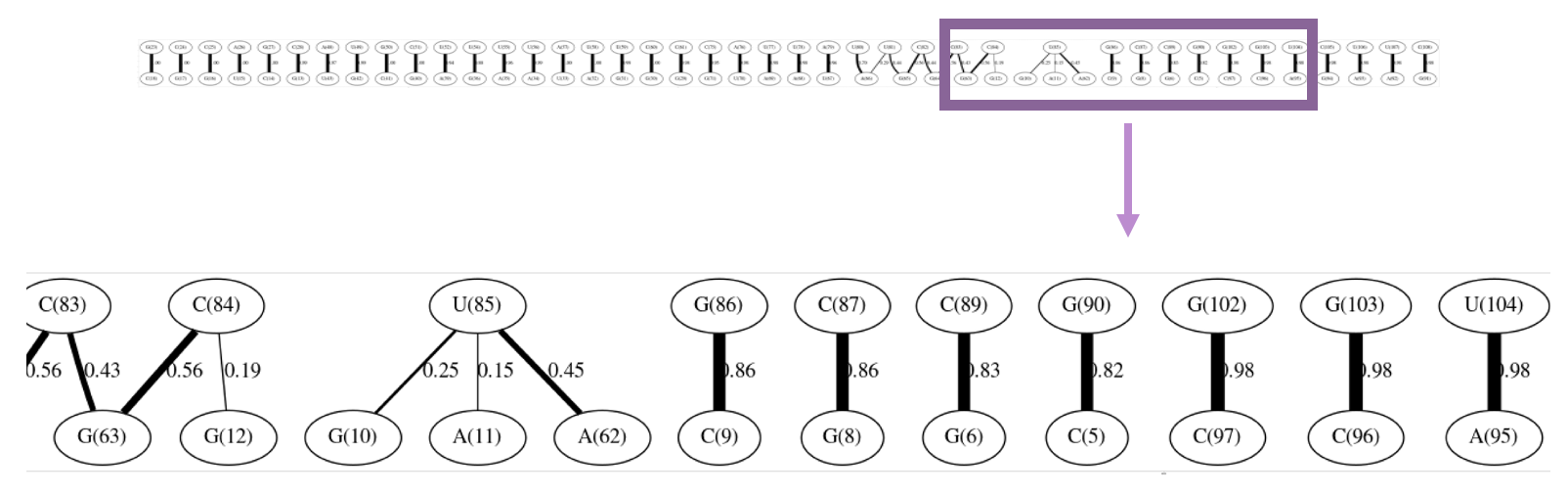}
    \caption{Visualization of a an excerpt from sample graph structure of a BPPM. Information in the nodes correspond to the type of base and its place on the sequence, the width and number next to the edges refer to the interaction between the two bases with values ranging from 0-1 (produced by author).}
    \label{fig:graph_vis}
\end{figure*}

\begin{figure*}[ht]
    \centering
    \includegraphics[scale=0.8]{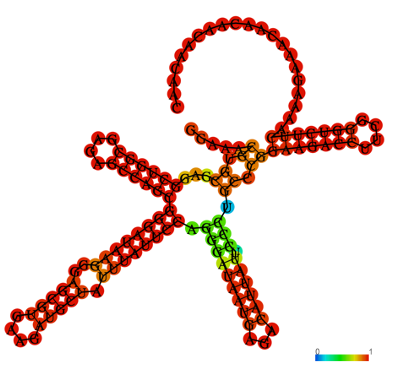}
    \caption{A `normal' visualization of a secondary mRNA nucleotide base sequence (produced by author).}
    \label{fig:normal_vis}
\end{figure*}

As mentioned earlier, each base of the sequence has five target and two further structural features associated with it. This will be represented as a feature matrix for all three ML models. What is of particular interest however, is the Base Pair Probability Matrix associated with each sequence which takes the form $N \times N$, where $N$ is the number of bases in the sequence. This probability matrix was generated through a computational calculation, optimizing for the lowest possible entropy configuration of the secondary structure. It can be represented as a standard matrix (refer to Figure ~\ref{fig:bppm} for visualization) or as a graph (refer to Figure ~\ref{fig:graph_vis}) with nodes and edges (compared to a `normal' visualization of such a secondary sequence - refer to Figure ~\ref{fig:normal_vis}). The higher the likelihood of two bases interacting, the higher the likelihood of a non-covalent interaction (hydrogen bonding) occuring between the two nucleotides. This distinction between representations is important for the former form is used for the two RNN architectures whereas the latter is used for the GCN. 

\subsection{Methods and Models}

\subsubsection{Data Processing and Overall Model Architecture}
Data from the dataset was first mapped to vector matrices and hot encoded, to ensure the data could be processed by a computer. Essentially, they were mapped onto a vector of the shape $1 \times n$ where $n$ is the number of possible options present for that specific feature. For example, since there are four nitrogenous bases, the vector shape to denote a single base was $1 \times 4$ and the total sequence vector for one strand of RNA took the shape $1 \times 4 \times 107$ (refer to Table ~\ref{tab:description} for information about the number of features).

After splitting the training and testing data according to the specifications lined out in Section 2.1.2, the training data was augmented using the ARNiE package, the same package used to create the data in the first place. Data augmentation is a technique to artificially create new training data from the existing data to improve the performance of ML models \cite{dataaug}. The  part of the package, Vienna2, that was originally used to predicts the RNA structures for the sequences, is based off of thermodynamic experiments and finds the structure by modelling the position of the lowest possible entropy. To augment the data, an alternate method was employed, CONTRAfold, that uses statistical methods to predict the structure. Therefore, for every sequence, new structures and predicted loop types were generated, effectively allowing for the training data to be doubled.

\begin{figure*}[ht]
    \centering
    \includegraphics[scale=0.28]{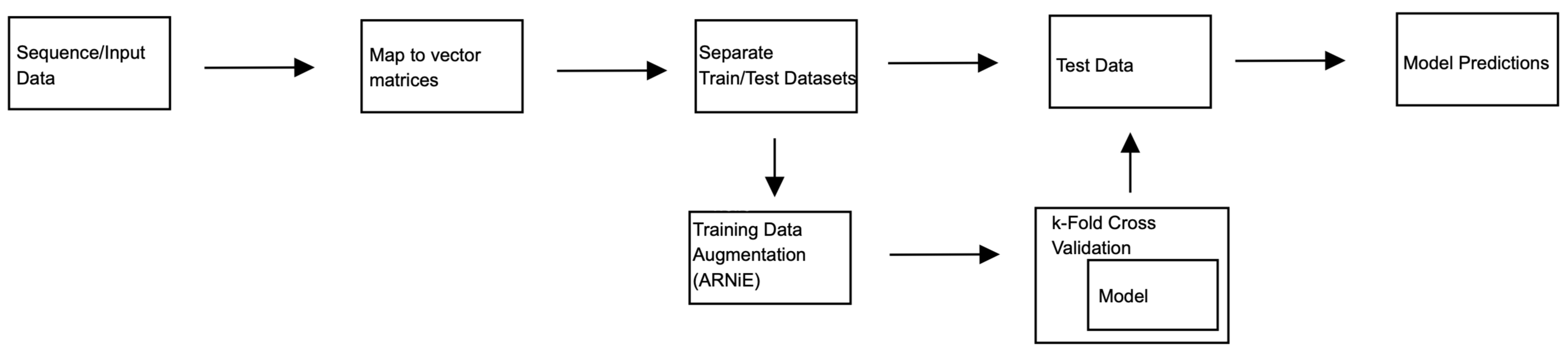}
    \caption{High level diagram of the overall model architecture. 'Model' in the diagram refers to one of the three ML approaches discussed in the next section (produced by author).}
    \label{fig:overall}
\end{figure*}

Finally, the resulting data was used to create the model using the architectures explained in Figure ~\ref{fig:overall} and the model was evaluated on the testing data. In terms of hyperparameter optimization, the number of epochs was set to 50 and the batch size was set to 64. K-Fold cross validation was also conducted with $k=4$. These values were determined through a simple grid search method to find the optimal quantity/quantity after which returns were negligible. An ADAM optimzer was also utilized with the learning rate set to a standard $1.0 \cdot 10^{-3}$, due to the fact that the model was running on a Tensorflow backend.

\subsubsection{Machine Learning, Deep Learning, and Neural Networks Background}
Although words such as `Artificial Intelligence', `Machine Learning', `Deep Learning', and `Neural Networks' have become popularized as buzzwords and seem to be synonymous with one another, there are important distinctions between them that must be understood.

Artificial Intelligence is the broadest term that denotes any computational technique that allows an algorithm to mimic human behavior by making decisions based on logic. It was first coined by John McCarthy in 1956 \cite{smith} and over the past 60 years fields within AI have undergone rapid advancement including but not limited to: searching algorithms, evolutionary computation, and robotics\cite{sortingreview, evoreview, roboticsreview}.

Machine Learning is one such subfield in AI that refers to the techniques that give computers the ability to learn and carry out tasks without explicitly programmed instructions. This usually entails a program performing statistical operations on a large dataset and discovering underlying patterns. For example, a Support Vector Machine (a type of ML algorithm) might be able to analyze labeled patient data then categorize unlabeled data into risk classes of a certain disease. 

Finally, Deep Learning is an even more specific term that refers to the subset of ML which employs multi-layer neural network algorithms to learn \cite{deeplearning}. For certain tasks, such as in Computer Vision or Natural Language Processing, the usage of such networks can recognize patterns that may be hidden to simpler forms of ML, therefore improving the performance of these algorithms.

Neural networks are computational models that are loosely based on the human brain. Neurons in the brain are connected to many other and are continuously receiving electrical impulses at their dendrites which eventually reach the cell body. Here, they are integrated in some way and, crudely speaking, once some threshold is reached, the neuron 'fires' and an impulse is transmitted to other neurons through a synapse. Nodes (or neurons) in an artificial neural network operate in a similar fashion. They receive data from a node in a previous layer, perform a learnable computation on them, then output the result to the next layer in the network. A general computation taking place in a single node follows the form:

\begin{equation}\label{eq:ann1}
	n_{l+1} = \sigma(W_l \cdot n_l + b_l)	
\end{equation}

\begin{figure}[H]
    \centering
    \includegraphics[scale=0.25]{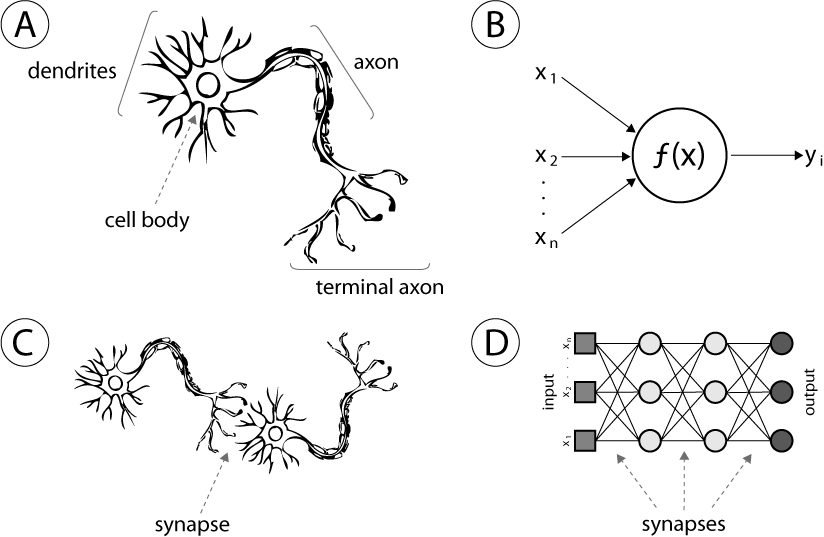}
    \caption{Comparison of neurons, nodes, synnapses, and network connections between cell bodies and an artificial network (Kim \cite{neuron_network}).}
    \label{fig:neuron}
\end{figure}

Where $n_{l+1}$ is the node which is being updated, $n_l$ represents all the previous nodes connected to $n_{l+1}$, $W_l$ are the weights, $b_l$ are the biases, and $\sigma$ is the activation function. These weights and biases are learnable, i.e. these are the values that through backpropagation and gradient descent, the computer is trying to optimize. The activation function is analogous to the binary threshold of the neuron in the brain to determine whether or not it will fire, however, in modern networks, more sophisiticated functions are used such as the sigmoid, ReLu, or hyperbolic tangent functions \cite{deeplearning, schmidhuber, emmert}. Figure ~\ref{fig:neuron} shows a visual comparison of biological and artificial neural networks.  

The type of neural networks that will be discussed in this paper are known as Recurrent Neural Networks (RNNs). They are, roughly speaking, repeating ANNs that allow for information through data in which spatial or temporal information is important \cite{rnnreview}. For example, an algorithm in a messaging app that has the aim of predicting the next word in the sentence not only needs information about the preceeding words in the sentence, but also the order in which they are located.

\subsubsection{Long Short Term Memory Networks}
A Long Short Term Memory Network (LSTM) is a type of RNN introduced by Hochreiter \&  Schmidhuber in 1997 \cite{hochreiter} and further popularized by many other pieces of work following it. Like other RNNs, it essentially consists of copies of the same network in which the output of the previous network (or cell) informs the next one, allowing information to persist. A single cell in an LSTM, consists of four distinct neural network layers, three of which are ones that make 'decisions' about the persistence of information as shown in equations ~\ref{lstm_1} -~\ref{lstm_3}:
\begin{equation}\label{lstm_1}
    f_t = \sigma(W_f[h_{t-1}, x_t] + b_f)
\end{equation}
\begin{equation}\label{lstm_2}
    i_t = \sigma(W_i[h_{t-1}, x_t] + b_i)
\end{equation}
\begin{equation}\label{lstm_3}
    o_t = \sigma(W_o[h_{t-1}, x_t] + b_o)
\end{equation}
Here the output of the previous cell along with the new input is passed through a basic neural layer with a non-linear activation function (traditionally sigmoid functions were used but in this paper, all activation functions denoted by $\sigma$ were ReLu). In parallel, candidate values for the cell state are also produced as shown by equation ~\ref{lstm_4}:
\begin{equation}\label{lstm_4}
    \tilde{C_t} = \phi_h(W_C[h_{t-1}, x_t] + b_C)
\end{equation}
Where the activation function in this case is a hyperbolic tangent function (compressing the values between -1 and 1). Finally using these output matrices, one can then update the cell state as given by equation ~\ref{lstm_5} and produce new outputs as shown in equation ~\ref{lstm_6}:
\begin{equation}\label{lstm_5}
    C_t = f_t \cdot C_{t-1} + \tilde{C_t} \cdot i_t
\end{equation}
\begin{equation}\label{lstm_6}
    h_t = o_t \cdot \phi_h(C_t)
\end{equation}
This output is then fed into the next cell. This model was used with the normal vector matrix BPPM as represented in Figure ~\ref{fig:bppm}. 
\subsubsection{Gated Recurrent Unit Networks}
Introduce by Cho et al. in 2014 \cite{cho}, a Gated Recurrent Unit Network is a variation of traditional LSTM networks that have the primary advantage of faster computation (due to less neural nets in each cell). Like in the LSTM, decision matrices are produced through two non-linear activation function based neural networks as shown in equations ~\ref{gru1} and ~\ref{gru2}:
\begin{equation}\label{gru1}
	r_t = \sigma(W_r[h_{t-1}, x_t] + b_r)
\end{equation}
\begin{equation}\label{gru2}
    z_t = \sigma(W_z[h_{t-1}, x_t] + b_z)
\end{equation}
Candidate values are also generated through a hyperbolic-tangent-based network, however, in this case it is done directly for the output as there is no cell state, unlike in an LSTM. The candidates are then chosen by the 'decision' matrices to produce the output as shown in equations ~\ref{gru3} and ~\ref{gru4}:
\begin{equation}\label{gru3}
   \tilde{h_t} = \phi_h(W_h[h_{t-1}, x_t] + b_h)
\end{equation}
\begin{equation}\label{gru4}
    h_t = (1-z_t) \cdot h_{t-1} + \tilde{h_t} \cdot z_t
\end{equation}
Like any other form of RNN, this output is then passed to the next hidden layer.
\subsubsection{Graph Convolutional Networks}
Graph Convolutional Networks were introduced by Kipf \& Welling in 2017 \cite{kipf} and provide a novel way to analyze arbitrarily structured data in the form of a graph. A GCN is not a form of an RNN although they are both connectionist models. A GCN operates on a graph defined by $G = (V, E)$ where $V$ is the set of nodes and $E$ the set of edges. Nodes in the graph aggregate the features of the surrounding nodes and itself and use the following neural net (refer to ~\ref{eq:gcn} to generate an output which is then assigned to the node:
\begin{equation}\label{eq:gcn}
	h_t = \sigma(W_h \cdot D^{-1}[h_{t-1}, \hat{A}] + b_h)
\end{equation}

Where $D$ is the diagonal node degree matrix of the graph and $\hat{A}$ is equal to $A + I$, $A$ being the adjacency matrix (taking the form $N \times N$) representing the graph (in this case, the BPPM - see Figure ~\ref{fig:graph_viz}). $h_0 = N \cdot F_0$ i.e. it is a feature matrix. After an output is generated, this process can be repeated each time the output of the nodes propagating outwards. Due to the localization of such a problem, only two repetitions were employed, any further increases resulted in negligible influences. 
\subsubsection{Performance Assessment}
The models will be evaluated based on the error produced in the prediction of the target values. The two loss measures employed in this paper are Mean Absolute Error (MAE) and Root Mean Square Error (RMSE) described in equations ~\ref{eq:mae} and ~\ref{eq:rmse}.
\begin{equation}\label{eq:mae}
    RMSE = \sqrt{\frac{\Sigma_{i=1}^n (\hat{y_i} - y_i)^2}{n}}
\end{equation}
\begin{equation}\label{eq:rmse}
    MAE = \frac{1}{n}\Sigma_{i=1}^n |\hat{y}_i - y_i|
\end{equation}

The difference between the two metrics is that in MAE, all the errors are averaged by weighting them equally, however, since RMSE has a quadratic term, larger individual errors will be punished more than smaller ones.
\section{Results}
\subsection{Initial Training and Validation}
After being built using the specifications discussed earlier in this manuscript, the models were trained on the `training' dataset discussed in section 2.1.2. For k-fold cross validation, in the models presented, 
\begin{figure*}[h]
     \centering
     \begin{subfigure}[b]{0.3\textwidth}
         \centering
         \includegraphics[width=\textwidth]{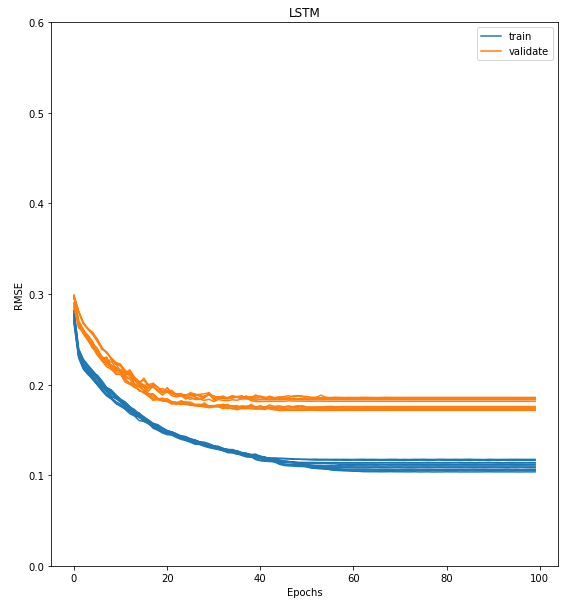}
         \caption{LSTM Model}
         \label{lstm_train}
     \end{subfigure}
     \hfill
     \begin{subfigure}[b]{0.3\textwidth}
         \centering
         \includegraphics[width=\textwidth]{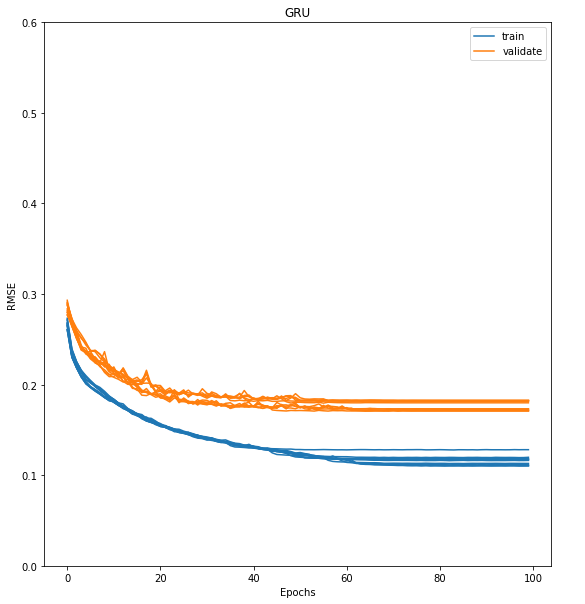}
         \caption{GRU Model}
         \label{gru_train}
     \end{subfigure}
     \hfill
     \begin{subfigure}[b]{0.3\textwidth}
         \centering
         \includegraphics[width=\textwidth]{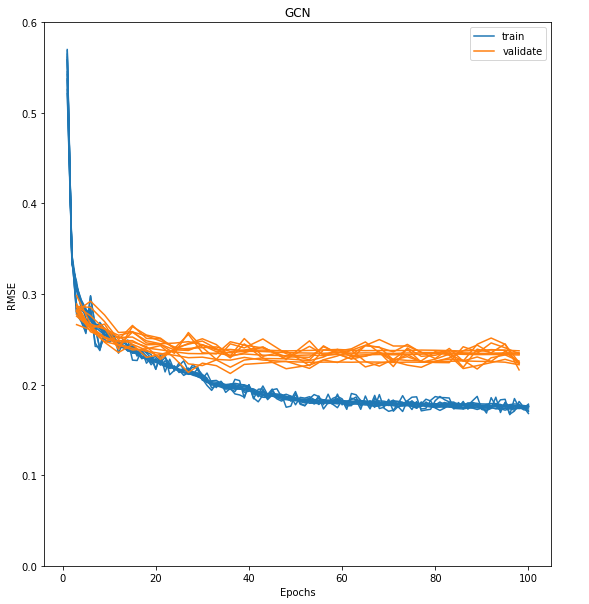}
         \caption{GCN Model}
         \label{gcn_train}
     \end{subfigure}
     \caption{Training and validation loss for the twelve histories of the ML Models (produced by author).}
        \label{training}
\end{figure*}
$k=4$ and this was initially repeated three times to improve accuracy. The model was split into four groups for cross validation as larger values of k were found to have negligible influence on the model results. Training and validation loss (RMSE) for the three models are show in Figure ~\ref{training} and Table ~\ref{tab:training}.

\begin{table}[h]
\centering
\begin{tabular}{lllll}
\Xhline{3\arrayrulewidth}
     & \multicolumn{2}{l}{Training} & \multicolumn{2}{l}{Validation} \\ \cline{2-5} 
     & RMSE              & MAE               & RMSE               & MAE                \\
     \Xhline{3\arrayrulewidth}
LSTM & 0.1089            & 0.0663            & 0.1758             & 0.1052             \\
GRU  & 0.1143            & 0.0693            & 0.1787             & 0.1072             \\
GCN  & 0.1752            & 0.1055            & 0.2232             & 0.1394             \\ \Xhline{3\arrayrulewidth}
\end{tabular}
\caption{\label{tab:training} Mean loss values of models after the training and validation processes.}
\end{table}

As can be seen after the initial training and validation phase, the LSTM maintains the best performance. The GRU scores were similar to that of the LSTM, however, the GCN was in a distant third place averaging about 0.05-0.06 higher RMSE values in comparison to the other two.

\subsection{Model Comparison}
\begin{figure}[H]
    \centering
    \includegraphics[scale=0.25]{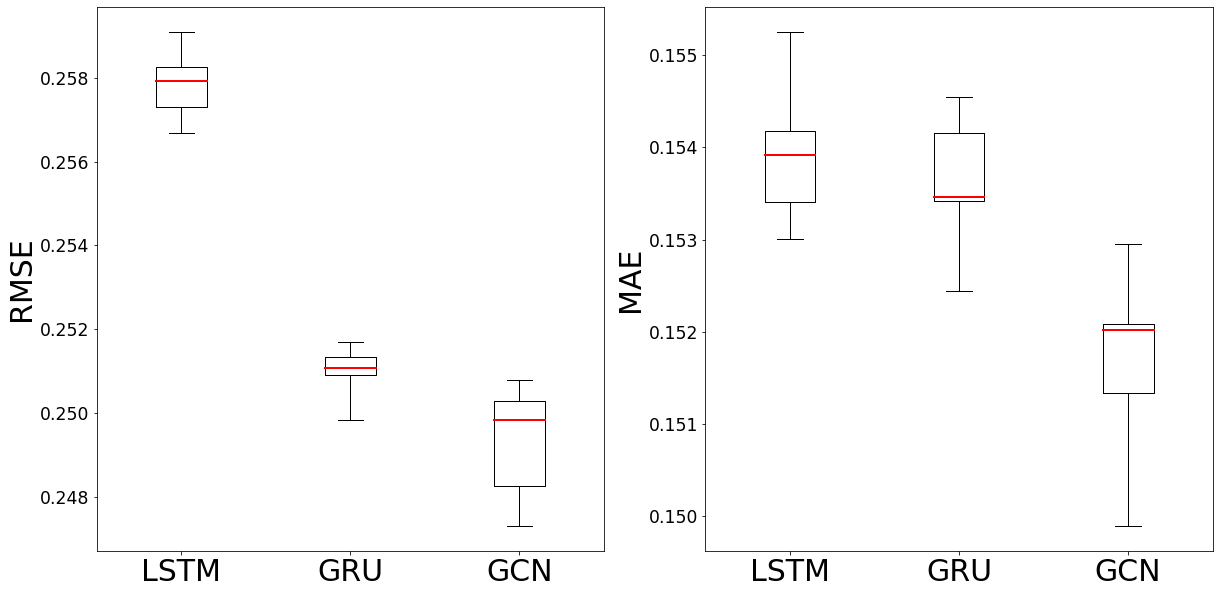}
    \caption{Results (loss) for over 20 trials of the models predicting the \textbf{reactivity} values of the training set sequences (note change in scale between plots). The GCN performed the best, both in terms of RMSE and MAE, followed by the GRU and LSTM respectively. Boxplots show the minimum, maximum, and median values along with the 25th and 75th quartiles (produced by author).}
    \label{fig:reactivity}
\end{figure}

\begin{figure}[H]
    \centering
    \includegraphics[scale=0.25]{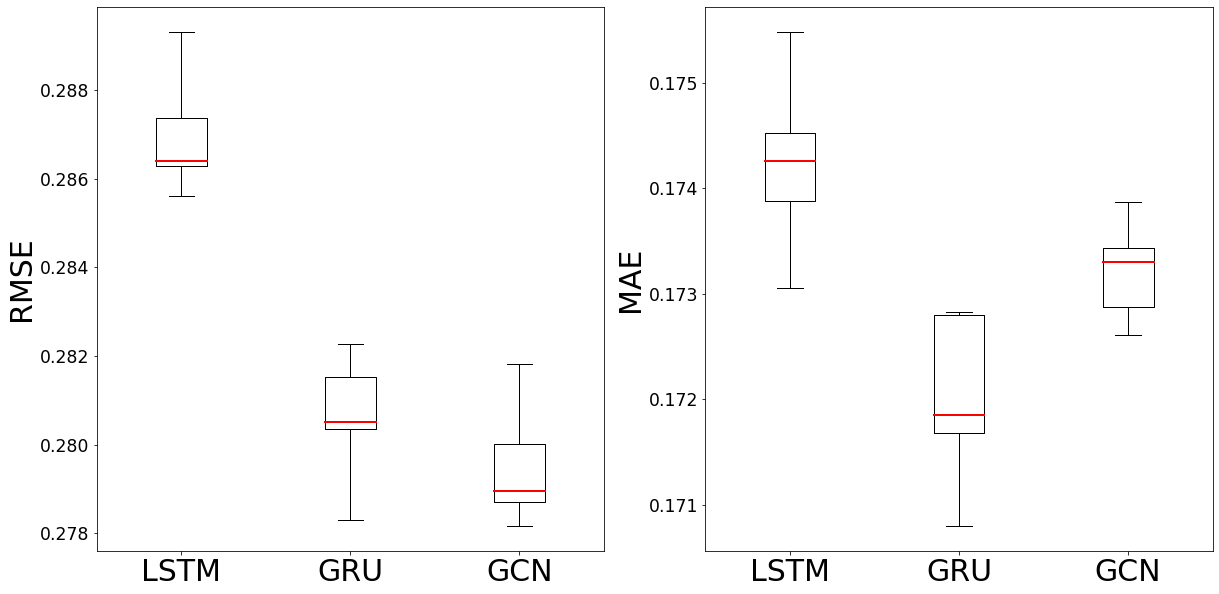}
    \caption{Results (loss) for over 20 trials of the models predicting the \textbf{degradation at pH 10} of the training set sequences (note change in scale between plots). The GCN performed the best in terms of RMSE, however, the GRU performed better in terms of MAE, suggesting the GRU is more prone to larger errors. Boxplots show the minimum, maximum, and median values along with the 25th and 75th quartiles (produced by author).}
    \label{fig:pH10}
\end{figure}

\begin{figure}[H]
    \centering
    \includegraphics[scale=0.25]{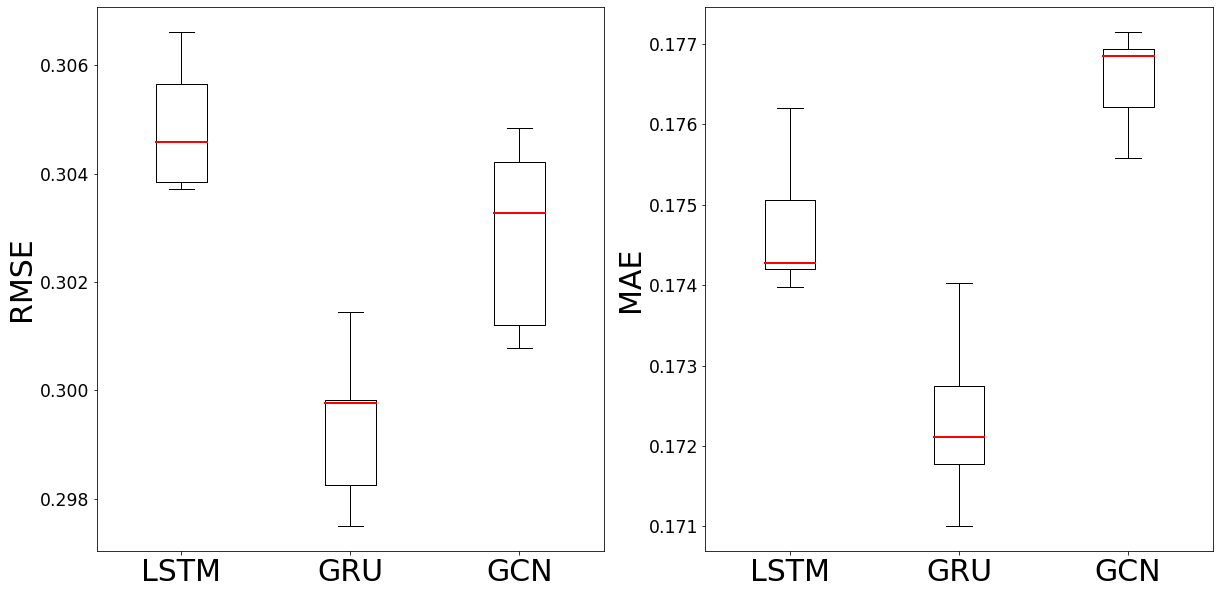}
    \caption{Results (loss) for over 20 trials of the models predicting the \textbf{degradation at pH 10 with added Magnesium} of the training set sequences (note change in scale between plots). The GRU performed the best, both in terms of RMSE and MAE, followed by the GCN in RMSE and LSTM in MAE, further reaffirming the notion that the GCN is less prone to larger individual errors. Boxplots show the minimum, maximum, and median values along with the 25th and 75th quartiles (produced by author).}
    \label{fig:pH10mg}
\end{figure}

\begin{figure}[H]
    \centering
    \includegraphics[scale=0.25]{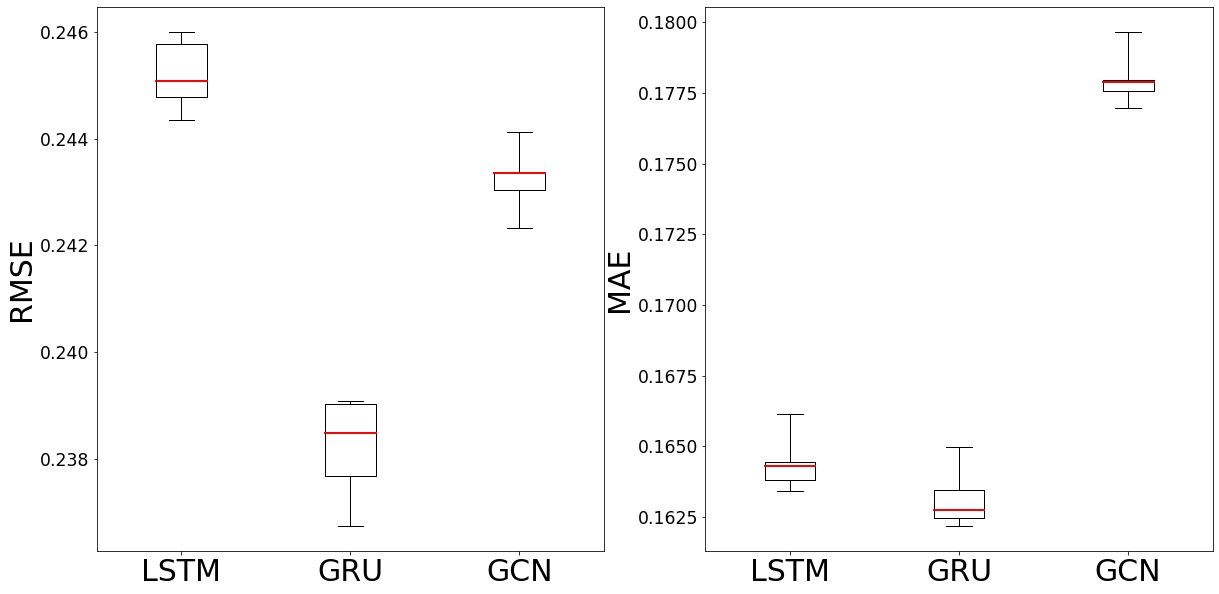}
    \caption{Results (loss) for over 20 trials of the models predicting the \textbf{degradation at 50$^\circ$C} of the training set sequences (note change in scale between plots). The GRU performed the best, both in terms of RMSE and MAE, the GCN in RMSE and LSTM in MAE, similar to Figure ~\ref{fig:pH10mg}. Boxplots show the minimum, maximum, and median values along with the 25th and 75th quartiles (produced by author).}
    \label{fig:50c}
\end{figure}

\begin{figure}[H]
    \centering
    \includegraphics[scale=0.25]{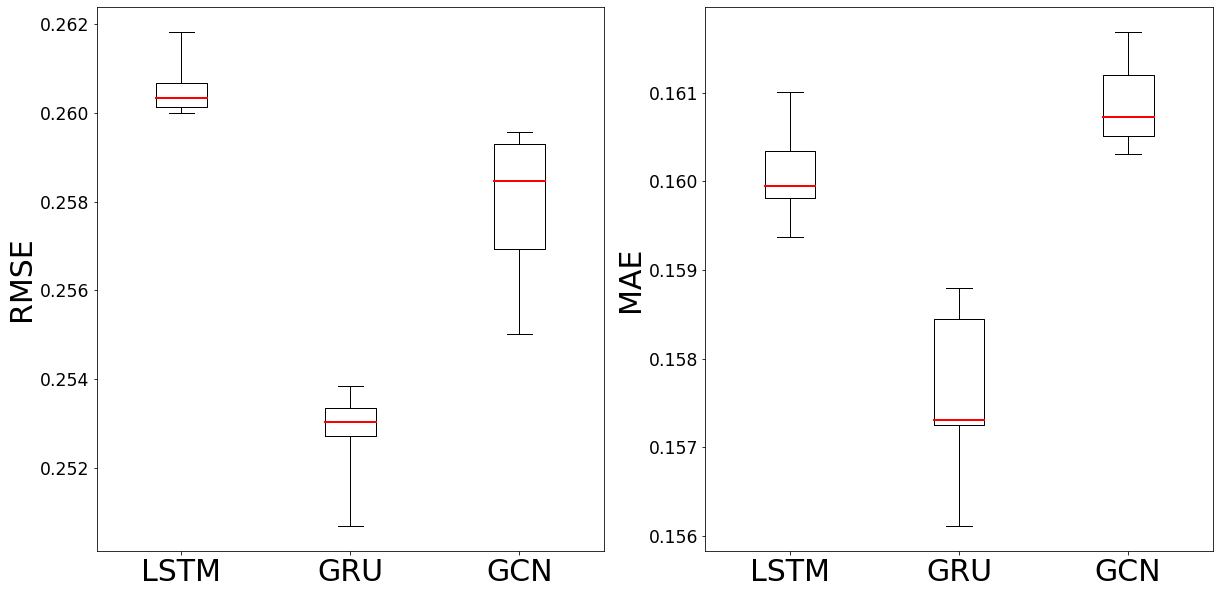}
    \caption{Results (loss) for over 20 trials of the models predicting the \textbf{degradation at 50$^\circ$C with added Magnesium} of the training set sequences (note change in scale between plots). The GRU performed the best, both in terms of RMSE and MAE, the GCN in RMSE and LSTM in MAE, similar to Figures ~\ref{fig:pH10mg} and ~\ref{fig:50c}. Boxplots show the minimum, maximum, and median values along with the 25th and 75th quartiles (produced by author).}
    \label{fig:50cmg}
\end{figure}

The performance for the three machine learning models on the training dataset are presented in Figures ~\ref{fig:reactivity} -~\ref{fig:50cmg}. The GRU maintained the lowest loss values for all target values, with the only exception being the reactivity (see Figure ~\ref{fig:reactivity}) and the degradation at pH 10 (only the RMSE - see Figure ~\ref{fig:pH10}) in which the GCN had the best predictions. Notably, in Figures ~\ref{fig:pH10mg}, ~\ref{fig:50c}, and ~\ref{fig:50cmg}, the GCN had lower RMSE values than the LSTM, however the opposite was true for the MAE values suggesting that the GCN is not as prone to large, individual errors.


\begin{figure*}[h]
    \centering
    \includegraphics[scale=0.15]{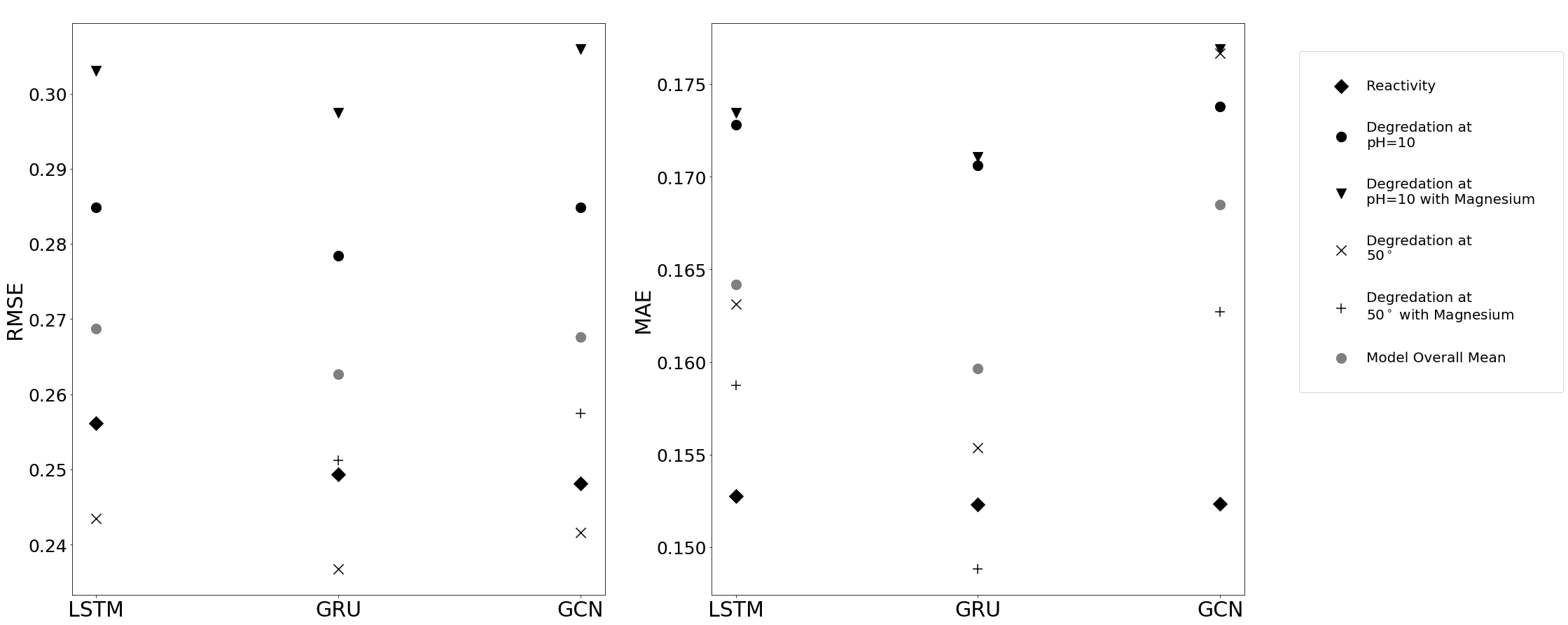}
    \caption{Mean loss results (RMSE and MAE - note change in scale between plots) for the three machine learning methods. The GRU performed the best across both metrics (produced by author).}
    \label{fig:scatter}
\end{figure*}

Mean loss across the 20 trials for each individual target were calculated along with that of all  targets combined for each model. Results are shown in Figure ~\ref{fig:scatter}.

\section{Discussion and Conclusion}
Although the creation of stable mRNA molecules remains difficult, datasets of sequences and corresponding information are becoming more popular and widely available. Through the use of deep learning architectures, reasonable predictions of structural features can be obtained, as demonstrated in this manuscript, with mean RMSE values ranging from 0.24 to 0.31. The usage of such techniques has the capacity to increase the speed and efficiency of mRNA vaccine discovery and has further implications in other related fields of research. 

However, such methods, as presented in this paper, are not without their limitations. It is important to note that while reasonably accurate results are produced, the error, when taking into account the scale of the values being predicted (0.5 - 4) is not insignificant; in fact, it equates to a 24\% Mean Absolute Percentage Error. Therefore, this could have the potential of incorrectly predicting the stability of an important molecule which may be overlooked during the discovery phase. Thus, in its current state, as an extension, a simple binary classification system is suggested to predict whether the molecule is stable, at the end of each of the models that would aim to minimize the False Negative rate - effectively resulting in the model being used as a screening test to remove highly unstable sequences, rather than a full-fledged research tool.

Another limitation of this method is the length of the sequences of the mRNA molecules used. The length used in this paper ranged from 107-130 bases, while an actual Covid-19 vaccine would likely range from 3000-4000 \cite{zhang} bases long. Thus, as more expansive and complete databases are released, further research exploring the reliability of such algorithms in predicting longer sequences is encouraged. 

Despite these limitations, the results of this work show that such prediction algorithms are feasible and have the potential to save time during research processes, an especially valuable commodity during disease outbreaks. In the long term, such techniques may also better help researches in understanding the reasoning behind the stability of certain RNA molecules and aid in the development of related technologies. It is hoped that this work will be of some use to other chemical researchers as well in creating better prediction models for this field.

\nocite{*}

\bibliographystyle{plainnat}
\bibliography{ref.bib}

\end{document}